\documentclass[11pt,preprint]{aastex}
\def\be{\begin{equation}}
\def\ee{\end{equation}}
\def\go{\mathrel{\raise.3ex\hbox{$>$}\mkern-14mu
             \lower0.6ex\hbox{$\sim$}}}
\def\lo{\mathrel{\raise.3ex\hbox{$<$}\mkern-14mu
             \lower0.6ex\hbox{$\sim$}}}

\begin{document}
\title{Comment on ``The Effect of Vacuum Polarization ... in Strongly
Magnetized Plasmas'' by Ozel (astro-ph/0203449)}
\author{Dong Lai and Wynn C.~G. Ho}
\affil{Center for Radiophysics and Space Research,
Department of Astronomy, Cornell University, Ithaca, NY 14853;~~
E-mail: dong, wynnho@astro.cornell.edu}

\begin{abstract}
We have recently shown that in a highly magnetized 
neutron star atmospheric plasma, vacuum polarization can 
induce resonant conversion of photon polarization modes
via a mechanism analogous to MSW neutrino oscillation. 
In a recent paper Ozel has dismissed this mode conversion effect
as ``mistakes''. Here we explain why our arguments/calculations
of this effect are correct.
\end{abstract}

\section{Introduction}

There are two polarization modes for photons (with 
energy $E$ much less than the electron cyclotron energy)
propagating in a strongly magnetized plasma: The extraordinary mode (X-mode) 
and the ordinary mode (O-mode). These two modes have very different 
opacities in the atmosphere of a magnetized neutron star
(NS). In a recent paper (Lai \& Ho 2002; hereafter LH02), we discussed 
the effect of resonant mode conversion due to vacuum polarization: 
A photon propagating outward in the NS atmosphere can 
convert from one polarization mode into another as it traverses the 
resonant density, $\rho_v\simeq
Y_e^{-1}\eta^{-2}(B/10^{14} G)^2(E/1~{\rm keV})^2$ g cm$^{-3}$, where $Y_e$ is
the electron fraction, and $\eta\sim 1$ is a slowly varying function of the
magnetic field $B$. This mode conversion is analogous to the 
Mikheyev-Smirnov-Wolfenstein mechanism for neutrino
oscillation and is effective for $E\go {\rm a~few~keV}$.
Because the two photon modes have vastly different opacities,
the vacuum-induced mode conversion can affect radiative 
transport and surface emission from strongly magnetized NSs.
In Ho \& Lai (2002) (hereafter HL02), this effect was explored further
in numerical models of magnetar atmospheres (see HL02 for references of
earlier works related to strong-field vacuum polarization and
its effects on radiative transport). 

In a recent paper (Ozel 2002; hereafter Ozel02), Ozel has dismissed
the mode conversion effect discussed our papers as ``mistakes''.
The purpose of this note is to clarify why our arguments/calculations
of this effect are correct. Indeed, there are a number of problems in Ozel02,
and many of its criticism and description of our work are incorrect or
inaccurate. We restrict our comments to problems in Ozel02 that are directly
related to our work\footnote{We have previously communicated to Ozel in 
private our comments described in this note, but the comments 
were rejected by her. We therefore feel it is necessary to put this note on
astro-ph.}.

\section{Problems with Ozel (2002)}

{\bf 1.} 
Throughout the paper (especially in the appendix), Ozel02 dismisses the
MSW-like mode conversion effect first discussed in our papers.
The arguments in Ozel02 are incorrect.
Here is how to understand the effect. Adiabatic evolution
of a state (whether it is quantum mechanical wavefunction or polarization
state of a classical EM wave) means that the evolution is continuous, with no
sudden change of the parameters that characterize the state (wavefunction
or eigenvalue). Thus the different ways of describing the modes
are of real significance. The mode eigenfunctions are characterized
by the polarization ellipticity $K$: Eq.~(2.27) of HL02 gives
$K_j=\beta\left[1+(-1)^j(1+r/\beta^2)^{1/2}\right]$
for the X-mode and O-mode ($j=1,2$), and eq.~(2.43) gives
$K_\pm=\beta\pm (\beta^2+r)^{1/2}$ for the plus-mode and minus-mode.
Obviously, $K_1$ and $K_2$ are discontinuous across
the vacuum resonance (note that $\beta=0$ at vacuum resonance, and
$\beta$ changes sign across the resonance),  while $K_+$ and $K_-$ 
are continuous (see Fig.1 of LH02). Similarly, the indices of refraction 
(the eigenvalues) $n_1$ and $n_2$ are discontinuous, while $n_+$ and $n_-$ are
continuous across the resonance (in quantum mechanics, this is called ``avoid
crossing"). Thus, under adiabatic condition (i.e., when the density
gradient of the medium is sufficiently gentle), the polarization state will
evolve along the $K_+$ or $K_-$ curve, rather than along the $K_1$ or $K_2$
curve. Since $K_+$ manifests as the O-mode at high density and as X-mode at
low density (see Fig.1 of LH02), we have adiabatic mode conversion across
the resonance. This conversion is not a matter of semantics,
but has true physical significance: For example, without this
conversion the mode opacity exhibits a spike at the vacuum resonance (see the
left panel of Fig.~3 of HL02) but with conversion the opacity has a
plateau-like transition (the right panel of Fig.3). As we show both
analytically (see sect.~4 of LH02, where we discuss photon
decoupling depth in the case of no mode conversion and the case with mode
conversion) and numerically (see HL02), this will lead to a genuine difference
in the radiative transfer and the emergent spectra.

Note that the adiabatic condition should not be confused with
the condition of ``Faraday depolarization''.
The adiabatic condition requires the background density to vary slowly (the
precise condition is derived in LH02), while Faraday depolarization
implies that a modal description of radiative transport is possible
(as first discussed by Gnedin \& Pavlov 1974 and repeated in sect.~2 of
Ozel02; the issue of Faraday depolarization was briefly discussed in sect.~2.4
of HL02, as well as in sect.~2.1 of Ho \& Lai 2001).
Adiabatic mode conversion exists even in a lossless medium
(with no absorption and scattering) for which Faraday depolarization is
always satisfied.  The MSW analogy is: Neutrinos are essentially
lossless in the sun, yet they can convert from one flavor to another.

Also, the adiabatic mode conversion should not be confused with mode
switching due to scattering; the latter was treated in Ozel02, in HL02,
and in many previous papers (see footnote 1 of LH02 and sect.~2.4 of HL02).

{\bf 2.} In addition to the conceptual problems in Ozel02
as discussed above, there are also several (less important) issues we 
would like to clarify in response to Ozel02: 

(1) When mode conversion is neglected, the photon
opacity exhibits a sharp feature at vacuum resonance, and in HL02 we adopted 
an approximate ``equal-grid'' numerical scheme to handle this feature.
Ozel02 has criticized our ``equal-grid'' scheme.
While it is certainly fine to criticize this, it should be noted that 
the limitation (including the one discussed in Ozel02)
and usefulness of the equal-grid method are fully
discussed in HL02 (sect.~5.1; this is also mentioned at the end of sect.~6.3,
after we comment on previous works). In fact, the subtlety in 
resolving the density-dependent, narrow vacuum resonance feature 
(and in particular the fact that the region very close to the resonance
contributes most to the optical depth) was first discussed in our
papers [see sect.~4 of LH02 (in particular footnote 2) and sect.~3,~4,~5.1 of
HL02] --- there was no hint that this subtlety was appreciated in previous
papers (including Ozel 2001; this is why we discussed in detail these issues in
our paper, including comparison with previous works in sect.~6.3).  

(2) We have emphasized several times in our paper 
[e.g., see HL02, sect.~1 after eq.~(1.2), the end of sect.~2.4, the second
to last paragraph of sect.~7] that a more rigorous treatment of the 
problem of radiative transfer in strong magnetic fields with vacuum
polarization requires one to go beyond the modal description
(as used in all NS atmosphere works so far) and to solve
the transport equations for the Stokes parameters (i.e., including 
the ``off-diagonal'' terms of the photon density matrix).

\section{Conclusion}
We stand by the main results of LH02 and HL02, although 
one should keep in mind the limitations 
(as discussed explicitly in 
sect.~7 of HL02). Strong-field vacuum polarization introduces several novel
features in the radiative transfer whose full, rigorous 
solution remains an important problem. Much work remains to be done 
in the area of highly magnetized NS atmospheres/surfaces in order to confront 
current and future observational data.

\acknowledgments
We thank Lars Hernquist and Ira Wasserman for useful comments.
This work is supported in part by NASA grant NAG 5-8484 and
an Alfred P. Sloan fellowship to D.L.


\end{document}